\documentclass[runningheads]{llncs}
\usepackage{microtype}
\usepackage{listings}
\usepackage{cite}
\usepackage[T1]{fontenc}
\usepackage{float}
\usepackage{booktabs}
\usepackage{graphicx}
\usepackage{multirow}
\usepackage[hidelinks]{hyperref}
\usepackage{url}

\begin{document}

\title{Assistance to Autonomy: A Systematic Literature Review of Agentic AI across the Software Development Life Cycle}

\titlerunning{Agentic AI Across the SDLC: A Systematic Review}

\author{Spyridon Alvanakis Apostolou\inst{1}\thanks{Corresponding author.} \and
Jan Bosch\inst{1,2} \and
Helena Holmström Olsson\inst{3}}

\institute{Chalmers University of Technology, Department of Computer Science and Engineering, Göteborg, Sweden \email{spyalv@chalmers.se} \and
Eindhoven University of Technology, Department of Mathematics and Computer Science, Eindhoven, Netherlands \email{jan.bosch@chalmers.se} \and
Malmö University, Department of Computer Science and Media Technology, Malmö, Sweden \email{helena.holmstrom.olsson@mau.se}}

\maketitle

\begin{abstract}
Agentic AI in software product development is increasingly adopted by organizations, yet the field lacks a consolidated synthesis of where adoption is mature, which architectural patterns dominate, and what limitations and coping mechanisms exist in industrial deployments. This systematic literature review addresses these gaps by establishing a body of knowledge as a starting point. Following Kitchenham guidelines, we queried four major research databases, obtaining over 1600 candidate publications. To handle this volume, we developed and validated a domain-agnostic multi-agent screening pipeline that extends prior LLM-assisted review tools by combining automatic metadata curation, inter-agent iterative dialogue, and conflict-resolution defaults that minimize false negatives. From the 92 manually verified primary studies, our thematic synthesis reveals that output verifiability is the primary enabler of agentic adoption: later software development life cycle phases, whose outputs are objectively evaluable through executable feedback, demonstrate the highest maturity and industrial presence, while earlier phases remain almost exclusively academic proofs-of-concept. We identify the Planner–Executor–Reviewer role specialization as the dominant architectural pattern, with the Reviewer agent implementing verifiability through executable feedback loops. Across all challenge categories, industrial mitigation strategies converge on confining agent actions to verifiable, bounded spaces. This study contributes a comprehensive characterization of the current literature on agentic systems in software product development, and a methodological contribution in the form of an AI-assisted tool to automate the screening phase in high-volume SLR domains.
\end{abstract}


\section{Introduction}
The landscape of Artificial Intelligence has evolved rapidly with the advent of Large Language Models (LLMs). The era of generative AI (GenAI) was defined by ``passive'' interactions, where models generate text in response to explicit human prompts. According to Wang et al.~\cite{wang_survey_2024}, from 2023 the field has been progressing towards agentic AI, which, unlike standard reactive LLMs, is characterized by systems that demonstrate capabilities such as planning, reasoning, use of external tools, self-refinement, and execution of multi-step workflows while requiring minimal human intervention. With this task-oriented behavior, agentic AI pushes the boundaries of traditional LLMs across a wide variety of domains~\cite{acharya_Agentic_2025, sapkota_ai_2026}.

This shift is particularly significant for Software Engineering (SWE). GenAI systems have been integrated into the Software Development Life Cycle (SDLC) as supportive assistants for tasks such as code completion, automated testing, and code generation. The emergence of agentic capabilities reframes the role of GenAI in SWE from assistance to autonomy~\cite{hosseini_role_2025}. Research is increasing rapidly, yet industrial adoption remains in its early stages: inherent LLM limitations, trust concerns, the lack of consolidated architectural patterns, and the rapid obsolescence of underlying models~\cite{raheem_Agentic_2025} stand as significant barriers to the integration of agentic systems into SDLC phases. As the volume of publications increases rapidly, a systematic synthesis of the current state-of-the-art is necessary. Consequently, this study conducts a Systematic Literature Review (SLR) to analyze and categorize the current state of agentic AI adoption and applications in the SDLC. This is further detailed in the following research questions (RQs):

\begin{enumerate}
    \item In which phases of the SDLC is agentic AI currently demonstrating the highest maturity and industrial adoption?
    \item What architectural patterns (e.g., multi-agent systems, tool use) are predominant in agentic AI implementations?
    \item What are the reported challenges in using agentic AI in industrial environments, and how have these limitations been addressed?
\end{enumerate}

The body of literature related to this domain is large and growing rapidly, making traditional manual filtering of publications extremely time-consuming. To address this, we developed and employed a multi-agent system using role-specific LLM agents (Assistant and Evaluator) to automate the screening process and efficiently identify the set of relevant publications.

The contribution of our study is two-fold: first, and most importantly, the results of the systematic literature review, which characterize the current state of agentic AI across the SDLC, while revealing output verifiability as the cross-cutting principle that connects phase maturity, dominant architectural patterns, and industrial mitigation strategies. Second, the multi-agent screening pipeline that extends prior LLM-assisted review tools with automatic metadata curation, inter-agent argumentative dialogue, and inclusion-biased conflict resolution, validated to achieve a low false-negative rate. The remainder of this paper is organized as follows: Section~\ref{BA} analyzes the background of the concept and the related work, Section~\ref{ME} describes the methodology, Sections~\ref{RE} and~\ref{TV} present the research findings, answers to the research questions, and threats to validity, Section~\ref{CO} presents the conclusions and future work, and Section~\ref{DC} provides data and code availability.

\section{Background}\label{BA}

\subsection{The Agentic Turn in SDLC}

GenAI was integrated into SDLC phases as an assistant tool through reactive code completion and generation. Tools such as GitHub Copilot, Amazon CodeWhisperer~\cite{yetistiren_evaluating_2023}, and Meta's CodeCompose~\cite{murali_ai-assisted_2024} demonstrated measurable gains in developer productivity and code generation~\cite{peng_impact_2023,becker_measuring_2025}. Nevertheless, the assistants are bounded, responding to explicit prompts without taking autonomous actions beyond immediate output implementations. 

As a part of progress, LLM agents extended this baseline by introducing tool integration and a sense-plan-act structure, enabling task-specific automation. However, individual agents remain narrow in scope and operate independently across tasks, lacking features that enhance autonomy ~\cite{sapkota_ai_2026}. Agentic AI goes further, with capabilities such as task decomposition, persistent memory, and orchestrated autonomy, these systems pursue complex, multi-step goals with minimal human intervention~\cite{sapkota_ai_2026,acharya_Agentic_2025,schneider_generative_2025}. Practically, this distinction means agentic systems can span entire SDLC phases rather than individual developer actions.

\subsection{Related Work}

Research activity in this field has grown substantially since 2023, when the first agentic frameworks for end-to-end development and role-specialized multi-agent pipelines began to emerge~\cite{wang_survey_2024}. Several recent secondary studies approach the topic from complementary angles. He et al.~\cite{he_llm-based_2025} examined 71 primary studies on LLM-based multi-agent systems across the SDLC. Their review is paired with case studies that surface capability limits, and it concludes with a two-phase research agenda focused on individual agent enhancement and, subsequently, agent synergy. A broader synthesis is offered by Liu et al.~\cite{liu_large_2025}, who survey 106 papers through a dual lens: an SWE perspective spanning individual tasks and end-to-end workflows, and an agent-centric view that analyzes planning, memory, perception, action, and collaboration. Wang et al.~\cite{wang_agents_2024} take a different direction. Working through 115 papers under a perception-memory-action framework, they flag open issues such as under-explored modalities, hallucination mitigation, and inefficiencies in multi-agent coordination. Otoum and Elkhalili~\cite{otoum_methods_2026} narrow the lens to method categories, namely autonomous coding, multi-agent systems, iterative refinement, and human-agent collaboration, classifying 61 studies along these axes while tracing their temporal evolution and pointing to an early-phase autonomy gap.

Looking beyond software-specific reviews, Bandi et al.~\cite{bandi_rise_2025} cover broader ground. Their analysis of 143 studies across multiple domains contributes concept-level disambiguation of agentic AI paradigms, a comparison of major LLM frameworks, and five architectural models accompanied by evaluation-metric taxonomies. Hosseini and Seilani~\cite{hosseini_role_2025} shift the discussion toward organizational strategy, proposing an adoption framework that addresses business needs, tool selection, and risk management. The practitioner viewpoint is captured by Akbar et al.~\cite{akbar_Agentic_2025}: drawing on interviews with 21 experts, they argue that agentic AI redefines traditional SDLC phase boundaries through continuous feedback loops and real-time optimization.

We systematically examine peer-reviewed publications on agentic AI across the SDLC, contrasting academic proof-of-concept (PoC) studies with industrially adopted implementations to resolve SDLC-phase maturity (RQ1), architectural patterns in relation to agent autonomy (RQ2), and reported challenges paired with their mitigation strategies in industrial context (RQ3). Critically, rather than organizing findings around agent capabilities or method categories in isolation, our synthesis surfaces output verifiability as the cross-cutting principle that connects phase distribution, architectural design, and industrial mitigation strategies. Beyond the SLR results, the study makes a methodological contribution of a domain-agnostic multi-agent screening pipeline that extends existing LLM-assisted review tools with automatic metadata curation, inter-agent argumentative dialogue, and inclusion-biased conflict resolution, applicable to high-volume SLRs beyond this domain.

\section{Methods}\label{ME}

\subsection{Research Design}\label{RD}
As the basis for this SLR, we followed the Kitchenham and Charters guidelines \cite{kitchenham_guidelines_2007}.  We consulted an initial list of databases that included: Google Scholar, Scopus, IEEE Xplore, ACM Digital Library, and ScienceDirect. That initial list was further refined to address reproducibility and search functionality issues. Google Scholar was excluded because of the time and location dependence of its results, in addition to an insufficient character limit \cite{gusenbauer_which_2020}.  ScienceDirect was excluded due to its very restrictive term and boolean limit in advanced search \cite{gusenbauer_which_2020}. The refined list, using the terminology defined in \cite{gusenbauer_which_2020}, consisted of the following libraries: 
\begin{itemize}
    \item IEEE Digital Library http://ieeexplore.ieee.org
    \item ACM Digital Library http://dl.acm.org
    \item SpringerLink https://link.springer.com/
    \item Scopus https://www.scopus.com
\end{itemize}

\subsection{Query}\label{Query}
In order to translate the study goal with the research questions to a query structure, we utilized the CIMO (Context, Intervention, Mechanism, Outcome) logic \cite{denyer_developing_2008}. The key data points extracted included:
\begin{itemize}
    \item Context: The specific problem domain or application area.
    \item Intervention: The techniques and methods used at the context.
    \item Mechanism: The underlying methods and tools employed.
    \item Outcome: The results achieved, including performance improvements.
\end{itemize}

The implemented query is the following: 
\begin{lstlisting}[
    basicstyle=\ttfamily\small, 
    breaklines=true,                 
    columns=fullflexible,
    % The magic happens here:
    literate={AND}{AND\\}{3} 
]
("software lifecycle" OR "software development" OR "software product") 
AND 
("agentic" OR "autonomous agent*" OR "goal-driven agent*" OR "LLM agent*"  OR ("multi-agent" AND "LLM")) 
AND 
("tool use" OR "self-reflection"  OR "task decomposition" OR "orchestration" OR "proactive" OR "goal-driven" OR "task oriented" OR "self-adapted") 
AND 
("automation" OR "accelerat*" OR "efficiency")
\end{lstlisting}

\subsection{Time Frame, Inclusion \& Exclusion Criteria}
Since the study focuses on agentic AI in SDLC, we limited our review to publications from 2023 onwards, aligning with the timeline identified by Wang et al. \cite{wang_survey_2024} who identifies the first agentic frameworks to develop, and thus initiate the transition from standard LLMs to agentic AI as a distinct phase beginning with the advanced reasoning capabilities of models released in early 2023. 
To avoid collecting unnecessary publications, we employed the timeframe criteria as a filter at the advanced search phase along with the query implementation to each database.  
Additionally, in order to ensure adherence to the review scope, the inclusion and exclusion criteria shown in Table \ref{tab:inclusion_criteria} were curated to be applied to the initial search results. These criteria were designed using the guidance in Kitchenham and Charters \cite{kitchenham_guidelines_2007} and Gough et al. \cite{gough_introduction_2017}. 

Along with the time frame filter, we implemented the filter of language during the raw data collection phase when filtered data resulted from the query into the diverse databases. 

 \begin{table}[h]
    \footnotesize
    \centering
    \caption{Inclusion Criteria}\label{tab:inclusion_criteria}
    \begin{tabular}{cp{0.88\columnwidth}}
        \midrule
        \textbf{ID} & \textbf{Criteria} \\
        \midrule
        1 & Publication date within the defined time frame. \\
        \midrule
        2 & A primary study, not a secondary or a tertiary study. \\
        \midrule
        3 & Published in a peer-reviewed journal, conference proceeding, or book chapter not in grey literature reports, blog posts, or non-academic publications. \\
        \midrule
        4 & Published in English. \\
        \midrule
        5 & Not a duplicate or a version of another included study. \\
        \midrule
        6 & Discusses LLM agents with particulars matching the agentic characteristic. \\
        \midrule
        7 & Focuses in software product development. \\
        \midrule
        8 & Study provides a Proof of Concept (not a vision study). \\
        \bottomrule
    \end{tabular}
\end{table} 

\subsection{Data Analysis Method} \label{DAM}

Agentic AI is a rapidly evolving domain in which the literature grows continuously, making manual review at scale impractical. This motivated the development of an agile LLM-based tool to reduce reviewer workload. Existing tools such as Rayyan \cite{ouzzani_rayyanweb_2016} and Abstrackr \cite{wallace_deploying_2012} rely on active learning to prioritize relevant studies. More recently, LatteReview \cite{rouzrokh_lattereview_2025} employs role-playing LLM agents with hierarchical screening and reasoning transparency. However, LatteReview requires users to manually script agent personas and scoring rules, and expects a complete dataset as input.

To further automate the process and reduce both prompt subjectivity and false negatives, we developed a model- and domain-agnostic six-step pipeline. It requires minimal reviewer input, raw publication metadata (title, DOI, abstract, etc.) and a single detailed prompt containing the research purpose, research questions, and selection criteria. The pipeline includes self-curation of missing abstracts, inter-agent classification discussions, and produces binary relevant/irrelevant labels with full transparency of decisions, reasoning, and agent dialogue.

\begin{figure}[t!]
    \centering   
    \includegraphics[width=\columnwidth]{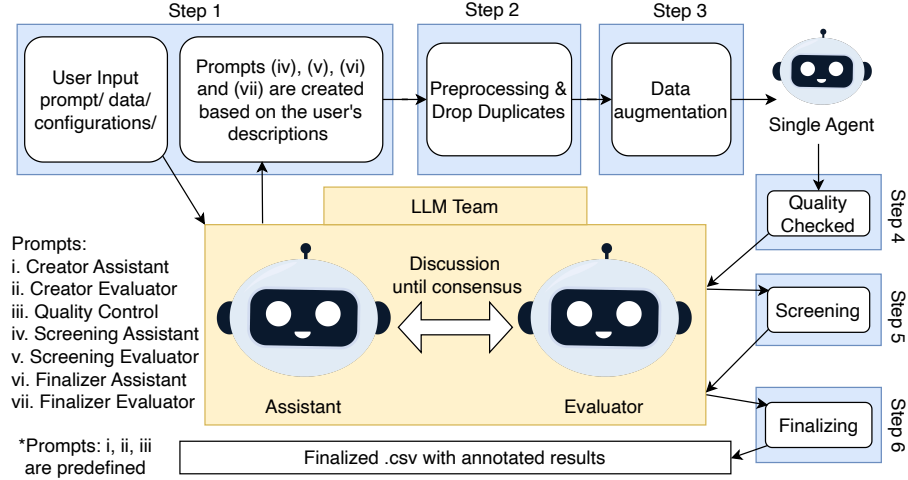}
    \caption{Analytical workflow of the multi-agent collaboration and consensus mechanism.}
    \label{fig:fullpipeline}
\end{figure}

In the first step of the workflow, we have two role-playing agents (Assistant and Evaluator) based on the user-provided prompt, respectively they generate and evaluate task-specific prompts for the subsequent phases, only after the reviewer approves the final prompt of each task the process proceeds to the next step. In the second step, raw exports from multiple databases (CSV, RIS, or BIB format) are normalized into a unified CSV with duplicates records removed. In the third step, a self-curation process retrieves missing abstracts via web scraping and API-driven augmentation.

From step four onward, records are filtered against the inclusion criteria. In the Quality Control step, a single agent excludes non-research items (e.g., conference preambles, generic book chapters). A human verification of excluded records lacking abstracts is recommended at this stage to prevent accidental omission of legitimate studies.

In the Screening step, the Assistant and Evaluator agents independently classify each publication against the inclusion/exclusion criteria and high-level domain relevance. It is important to note here that, for every double-agent step, we employed different models in order to minimize classification errors when independent reasoning paths reach the same conclusion. In case of conflict, a structured inter-agent dialogue with iterative argumentation (up to three rounds or until agreement) is triggered to finalize the classification label, unresolved conflicts default to inclusion to minimize false negatives. The purpose of this dialogue is to leverage one agent's reasoning to challenge the other's classification, serving as an additional de-hallucination mechanism. The final Relevance Selection step applies the same multi-agent process, now focused on alignment with the research questions and scope. The complete workflow is illustrated in Figure \ref{fig:fullpipeline}.

The two-phase (screening and relevance) design serves three purposes: (1) modularity, allowing costlier models to be selectively applied to the smaller candidate set in the final phase only, (2) reduced agent cognitive load, as each phase operates under a constrained set of criteria, and (3) independent evaluability of each stage.

Following the automated pipeline, the candidate relevant publications underwent manual evaluation following a structured two-pass protocol. In the first pass, each selected as relevant study's abstract was first read and applied Criteria 6, 7, and 8 from Table~\ref{tab:inclusion_criteria} to confirm or reject the automated label. In the second pass, studies confirmed as relevant were read in full to extract responses to at least one of the research questions. Data extraction was guided by a pre-defined coding sheet capturing: (i) targeted SDLC phase, (ii) evaluation context (academic benchmark vs.\ industrial deployment), (iii) architectural pattern, and (iv) reported limitations and mitigations.

\section{Results}\label{RE}

\subsection{Query and Data Analysis Implementation}

The search was executed across the databases and search systems described in Section \ref{RD}, using the query as described in Section \ref{Query}. Table \ref{tab:search_results} summarizes the results acquired from each phase of the process.

We initially provided the prompt-creation agents with a definition and a detailed description of the agentic AI concept, together with all the research questions and the exclusion criteria (except the language and time-range filters, which had already been applied at the database query stage). Based on these inputs, four prompts were automatically generated for the screening and relevance phases. We employed bigger models for the prompt-generation phase (Assistant (A): gpt-5.4, Evaluator (E): gpt-5.2) to ensure high-quality prompt generation, and smaller yet capable models for the subsequent phases to manage cost (A-scr: gpt-5-mini, E-scr: gpt-5-nano, A-rel: gpt-5-mini, E-rel: deepseek-v3).

\begin{table}[htbp]
    \footnotesize
    \caption{Search Process Results}
    \label{tab:search_results}
    \renewcommand{\arraystretch}{1.3}
    \begin{center}
        \begin{tabular}{|l|c|c|c|c|c|c|c|}
            \hline
            \textbf{Source} & \textbf{Raw} & \textbf{Proces.} & \textbf{QC} & \textbf{SC} & \textbf{Relev.} & \textbf{Manual Eval.} \\
            \hline
            ACM      & 659  & 605  & 139    & 44  & 16 & 9   \\ 
            \hline
            IEEE     & 289  & 284  & 247    & 106 & 55  & 41  \\
            \hline
            Scopus   & 485  & 326  & 315    & 88  & 43 & 32  \\
            \hline
            Springer & 276  & 116  & 95     & 27  & 13  & 10 \\
            \hline
            \textbf{Total} & \textbf{1609} & \textbf{1331} & \textbf{796} & \textbf{265} & \textbf{127} & \textbf{92}\\ 
            \hline
        \end{tabular}
    \end{center}
\end{table}

As illustrated in Table~\ref{tab:search_results}, a significant number of records corresponded to generic conference introductions rather than actual research publications and were consequently excluded. After completing the quality-control step, the final number of valid publications was 796. A manual inspection of the generated annotations from this step revealed that only 3 out of the 1331 retrieved records were still missing abstracts while still representing legitimate research publications.

We applied the final two steps of the process. It is worth noting that the screening phase is the most time-consuming step, as a large number of quality checked publication's abstracts must be evaluated by two independent agents. Out of 796 valid publications, 265 passed the screening phase, the remaining 531 were excluded for failing to meet the inclusion criteria beyond those already enforced (date range, publication type, and language, Criteria 1, 3, and 4 in Table~\ref{tab:inclusion_criteria}). The final relevance-selection step identified 127 candidate studies. Following the manual evaluation procedure described in Section~\ref{DAM}, which involved reading the abstracts of all 127 papers, we identified 101 as conceptually relevant, from which we read the full text. Finally, 92 were confirmed as the final relevant set, since 9 papers lacked a PoC and were excluded. Consequently, we identified 26 as false positives produced by the automated pipeline, since only the abstracts were provided to the agents. In addition to this manual verification, a random sample drawn from the quality-control and screening exclusion sets was evaluated to further validate the selection process and ensure that false-negative classifications were minimized.

\subsubsection{False negatives evaluation}

To evaluate the pipeline's reliability, we sampled 100 publications that were excluded by the agent-based stages: 50 excluded during the screening phase and 50 excluded during the final relevance selection phase. Following the central limit theorem rule of thumb, a minimum of 30 samples per stage was considered sufficient to approximate normality in the sampling distribution, and this is why we consider the samples of 50 sufficient. Manual evaluation of each sample was performed to identify any false negatives.

The single false negative identified across the full sample of 100 came from the relevance-selection phase (1/50), while no misclassifications were observed in the screening step (0/50). Extrapolating the observed rate across the full excluded population (669 papers), suggests approximately 7 missed relevant publications in total, and since the sole missed publication originated from the relevance-selection phase, which operates on a smaller candidate set, the true number is likely even lower.

\subsection{Thematic Synthesis}

Combining the information from the final set of 92 publications, we structured the answers to our initial research questions.  

\subsubsection{\textbf{RQ1}} In which phases of the SDLC is agentic AI currently demonstrating the highest maturity and industrial adoption?

The phase distribution of the final publications set is summarized in Table~\ref{tab:sdlc_distribution}. From 92 studies only 13 were developed and evaluated in an industrial context, while the rest 79 represent academic PoC studies evaluated on benchmarks or controlled experimental settings. For both types of studies we can observe from Table~\ref{tab:sdlc_distribution} that the phases of the SDLC that demonstrate higher representation are  \textbf{Testing \& QA}, \textbf{Maintenance}, \textbf{Deployment}, and \textbf{Coding \& Implementation} indicating that agentic AI maturity is concentrated in post-development phases. 

The dominance of post-development phases reflects an important characteristic. They include tasks from which the performance can be verified through objective, executable feedback (e.g., test results, compiler outputs, fault localization traces, and log signals). These concrete results provide the agents a specific state to improve during self-refinement loops, which can be implemented without human evaluation.

On the industrial side, the concentration in Testing \& QA reflects the state, that testing is a time consuming and labour intensive phase of the SDLC, and its outcomes are verifiable. Publications in this concept target high-cost bottlenecks with clear success criteria, from GUI testing of production desktop and web applications to automotive software release gatekeeping~\cite{khoee_gonogo_2024} and structural testing of agentic systems themselves~\cite{kohl_automated_2026}. Moreover, Deployment \& Operations tasks follow the same logic, as AIOps agents monitor structured operational signals (e.g., metrics, logs, and CI/CD states), which outputs can be evaluated against concrete production thresholds for service performance anomaly detection~\cite{ji_lemad_2025, khoee_gonogo_2024}. 

The underrepresentation of Coding \& Implementation and Maintenance with industrial implementation does not necessarily translate as lack of research effort in these areas. Academic studies classified under these phases demonstrate that the majority rely on test execution as primary verification mechanism. Studies targeting code generation, automated debugging, and program repair (corresponding examples of systems: EvoMAC, SoapFL, and RGD), validate their contributions through unit tests or benchmarks such as HumanEval or Defects4J ~\cite{hu_self-evolving_2024,qin_soapfl_2025,jin_rgd_2024}. Notably, Maintenance, Coding \& Implementation from the literature are not fully distinguished by the Testing \& QA phase, since the correctness of the agents outputs is eventually evaluated by a form of testing. This suggests that Testing \& QA represents an entry point for industrial agentic adoption, while the integration of agentic systems in Maintenance or Coding \& Implementation require agents to operate on large industrial codebases with legacy systems, undocumented constraints, and evolving requirements that controlled benchmarks do not capture.

\begin{table}[htbp]
\footnotesize
\caption{Distribution of Primary Studies by SDLC Phase and Evaluation Context}
\label{tab:sdlc_distribution}
\renewcommand{\arraystretch}{1.3}
\begin{center}
\begin{tabular}{|l|c|c|c|}
\hline
\textbf{SDLC Phase} & \textbf{Industrial} & \textbf{Academic} & \textbf{Total} \\
\hline
Maintenance                  & 1& 19& 20 \\
\hline
Testing \& QA                & 5 & 13 & 18 \\
\hline
Cross-cutting (multiple phases) & 4 & 11 & 15 \\
\hline
Deployment \& Operations     & 2 & 12 & 14 \\
\hline
Coding \& Implementation     & 0 & 12 & 12 \\
\hline
Requirements Analysis        & 0 &  7 &  7 \\
\hline
Project Management           & 1 &  2 &  3 \\
\hline
Design \& Architecture       & 1 &  2 &  3 \\
\hline
\textbf{Total}               & \textbf{13} & \textbf{79} & \textbf{92} \\
\hline
\end{tabular}
\end{center}
\end{table}
Finally, Requirements Analysis, Design \& Architecture, and Project Management represent the least explored phases, almost exclusively in academic settings. While limited in volume, the industrial context studies in Design \& Architecture and Project Management demonstrate that meaningful agentic integration is feasible even in early SDLC stages, ranging from formal architectural verification~\cite{tahat2025agree} to agile sprint planning automation~\cite{adapa2025multi}.

\subsubsection{\textbf{RQ2}} What architectural patterns (e.g., multi-agent systems, tool use) are predominant in agentic AI implementations?

The predominant architecture across the selected studies is multi-agent role specialization. In this concept complex SWE tasks are decomposed in sequential or graph-based pipelines of agents, each assigned with a specific and bounded responsibility. The most common structure is Planner → Executor → Reviewer, with an Orchestrator agent managing the flow of the sub-processes. Notably, this structure pattern appeared in all the SDLC phases publications. Moreover, the rationale is to reduce the cognitive load of the agents by narrowing  their scope in order to produce more predictable and accurate outcomes, while additionally structured typed inter-agent communication methods (e.g., via JSON or LangGraph state graphs) in many occasions replaced free text handoffs with formally typed state transitions. A key example of an industrial case study, illustrates why this structure succeeds in practice. A two-agent Planner–Actor configuration is used for software release gatekeeping~\cite{khoee_gonogo_2024}, where the Planner applies Chain-of-Thought reasoning with self-consistency to generate a release strategy, and the Actor executes from a predefined atomic action vocabulary with self-reflection for error correction. Restricting the Actor to a fixed action set was key to industrial adoption, as it bounds operational risk without sacrificing the reasoning flexibility of the Planner, achieving a separation between what to do and how to do it safely.

Beyond agent role decomposition, memory and retrieval architectures form the second structural axis of the surveyed systems. Memory and information retrieval with Retrieval-Augmented Generation (RAG) is implemented as a baseline mechanism for injecting domain-relevant context. Interestingly, we mostly find variations in industrial studies. For an agentic RAG system (validated in production settings at Apple)~\cite{hariharan_agentic_2025}, the authors developed a hybrid vector-graph knowledge system to preserve structural relationships between test artifacts and requirements, addressing this way the loss of relational context in flat retrieval. Similarly, the Codebase-Aware Generative Agents system~\cite{mavani_codebase_2026} uses a Zero-Shot Dependency Mapping Knowledge Graph as a shared memory layer across agents. By combining deterministic static code parsing with constrained LLM extraction, this approach enables reasoning over structural code relationships, addressing architectural blindness.

\subsubsection{\textbf{RQ3}} What are the reported challenges in using agentic AI in industrial environments, and how have these limitations been addressed?

This research question focuses on industrial studies, in which agentic systems are developed and evaluated under real-world company-specific contexts. Before we dive into it, it is important to note that there is a difference on perspective between academic and industrial based studies, nevertheless the limitations are similar. Academic implementations mostly focus on performance and feasibility under controlled experimental conditions, while industrial case studies attempt to address reliability issues to ensure consistency. Overall, the most common limitations we found in studies are the following: non-deterministic behavior, information overload, hallucinations, and scalability.

\textbf{Non-Determinism and Hallucinations:}  
Non-determinism and hallucinations in LLM outputs are universally acknowledged. In most studies (from both categories) the predominant coping mechanism is iterative self-correction loops, in which systems identify their own mistakes based on results from the evaluation method (e.g., testing, compiler output, logs, metrics). An extra feature to narrow the scope of agents is role assignments with  detailed instructions or chain of thought to provide information about steps, rather than letting the agents finding them on their own. Industrial studies, while adopting these coping mechanisms, implement additional constraint layers. More specifically, systems developed for domain-specific information implement a terminology knowledge base with semantic meanings to enhance the understanding of the agents, additionally in the same study, the complexity of the two-agent system were mostly assigned to one agent for transparency and interpretation reasons, while the second agent had a pre-defined action space \cite{khoee_gonogo_2024}. The latter is widely employed in industrial studies with variations (tool calling or API communication) or structured outputs of agents in order to reduce hallucinations from unstructured text. Notably, in a different perspective industrial study instead of trying to force the AI to be perfectly predictable, built a rigorous, automated safety net around the AI to catch it if it makes a mistake~\cite{kohl_automated_2026}, representing a shift that agents are not only tools but also subjects of verification.

\textbf{Scalability:}
Agentic AI systems in industrial contexts mostly operate on large codebases, distributed microservices, and continuous streams of logs. More specifically, scalability concerns in industrial contexts span computational cost, execution time, and context size. Multi-agent iterative loops and tree-search strategies (e.g., Monte Carlo exploration) introduce costs that grow non-linearly with task complexity. While in most studies task-decomposition is enough narrow the scope of the agent, cases that manage real time monitoring needs extra steps. For example, systems employed for continuous and massive-scale operations utilize a combination of decentralized execution with task delegation in containerized Kubernetes environments, and dynamic stream processing via engines that compute metrics over sliding time windows of data~\cite{ji_lemad_2025}. Other common solutions, target stage-specific pass/fail tests to reject intermediate errors directly rather than propagate them to later pipeline steps.

\textbf{Context Management and Information Overload:}
In industrial environments, the information stream is heterogeneous and continuous, as context is imported from source code, execution logs, documentation, and architectural artifacts. This limitation is not independent of the aforementioned ones, since in many occasions large information that exceeds the context window increases the likelihood of hallucinations or incomplete answers, while additionally requires more  computational time increasing the latency and the demand of sources to avoid bottlenecks. As described in RQ2, a common mitigation strategy in both industrial and non-industrial studies is the use of RAG to address this challenge. However, specifically in industrial contexts, a flat vector retrieval alone is unable to preserve structural relationships between artifacts. The CA-SAF system \cite{mavani_codebase_2026} addresses this with a Zero-Shot Dependency Mapping Knowledge Graph, replacing flat retrieval with graph-based reasoning over structural code relationships. Similarly, another testing system uses a hybrid vector-graph representation that preserves business-logic relationships between test artifacts and requirements \cite{hariharan_agentic_2025}. These approaches represent architectures where retrieval is structurally grounded rather than semantically approximate.

\subsection{Synthesis and Discussion}

Examining the three RQs jointly reveals a unifying principle: \textit{output verifiability} operates as the primary enabler of agentic adoption, and its influence is traceable across phase distribution, architectural design, and industrial mitigation strategy.

RQ1 shows that the phases with the highest maturity and industrial presence (Testing \& QA, Deployment \& Operations, and Maintenance) are those whose outputs are objectively evaluable via executable feedback. Requirements Analysis and Design \& Architecture, remain almost exclusively in academic PoC territory. This asymmetry suggests that the current ceiling of industrial agentic adoption is defined by the availability of ground-truth feedback at the task level.

RQ2 highlights this at the architectural level. The dominant Planner-Executor-Reviewer pattern is not only a coordination structure, the Reviewer agent is the verifiability mechanism, providing the feedback that grounds iterative refinement. The industrial preference for hybrid vector-graph retrieval over flat RAG adds a structural layer for verifiability to the information-access layer.  Finally, RQ3 shows that industrial mitigations across all challenge categories converge on confining agent actions to bounded, verifiable spaces (predefined actions, structured outputs, stage-wise pass/fail tests).

These observations imply that extending agentic integration into early SDLC phases is, first and foremost, a problem of designing phase-appropriate feedback mechanisms rather than of improving model capability.

\section{Threats to Validity}\label{TV}
\textbf{Threats to Internal Validity:} The search strategy was systematically defined and implemented across multiple major databases, although relevant studies may have been omitted due to query keyword limitations or database coverage. The automated pipeline was designed to minimize false negatives, to validate this we sampled excluded publications from two distinct phases of the six-step process and projected the estimated false-negative rate ($\approx$1\%) to the larger excluded set, estimating $\approx$7 misclassified studies. Furthermore, relevance decisions depend on predefined prompts and specific LLM models, meaning prompt inaccuracies or future model changes may introduce systematic variability in classification outcomes that cannot be fully controlled.

\textbf{Threats to External Validity:} This review focuses on peer-reviewed publications from 2023 onwards. Grey literature was deliberately excluded, as peer-reviewed studies provide methodological transparency and independent validation, however this entails an information loss since industrial agentic AI practices frequently emerge in grey literature prior to formal publication. Additionally, many industrial studies rely on proprietary data and organization-specific workflows, limiting the transferability of reported architectures and mitigation strategies across contexts.

\textbf{Threats to Temporal Validity:} Given the rapid evolution of agentic AI, the findings reflect the literature available at the time of data collection. However, the identification of open challenges and research gaps provides a forward-looking perspective that partially mitigates this limitation.

\section{Conclusions and Future Work}\label{CO}

This systematic literature review synthesizes the current state of agentic AI in software product development based on 92 primary studies identified through a structured search across four major scientific databases. The screening process was accelerated by a multi-agent pipeline that extends existing LLM-assisted review approaches with automatic metadata curation, inter-agent dialogue, and inclusion-biased conflict resolution, validated to achieve a low false negative rate.

The thematic synthesis across the three research questions reveals a unifying principle: output verifiability\ is the primary enabler of agentic AI adoption. RQ1 shows that the phases with the highest industrial presence (Testing \& QA, Deployment \& Operations, and Maintenance) are those whose outputs are objectively evaluable through executable feedback (test results, compiler outputs, operational metrics). Earlier lifecycle phases remain almost exclusively as purely academic PoCs, constrained by the absence of ground-truth feedback signals. RQ2 identifies the Planner-Executor-Reviewer role specialization as the dominant architecture, where the Reviewer implements verifiability within iterative refinement loops. RQ3 shows that across reported challenge categories (non-determinism, hallucinations, scalability, and context overload), industrial mitigation strategies consistently converge on confining agent actions to verifiable, bounded spaces.

\textbf{Future Work:} The most consequential open direction is extending agentic integration into other SDLC phases, which requires establishing phase-appropriate verifiability mechanisms to enable the iterative feedback loops that currently drive adoption in post-development stages. A second direction concerns the long-term assessment of production-scale agentic systems, evaluating their maintainability, adaptability, and reliability under evolving codebases and shifting requirements, conditions that controlled benchmarks do not capture. Finally, establishing empirical comparison protocols between multi-agent architectures across SDLC phases would allow practitioners to make principled decisions about agent autonomy and orchestration design.

\section{Data and Code Availability}\label{DC}
To facilitate the reproducibility of our methodology, we have made the complete replication developed tool available in an online repository.
\footnote{\url{https://doi.org/10.5281/zenodo.20670298}}. This repository contains the source code of the developed multi-agent pipeline, the specific prompts utilized for the Assistant and Evaluator agents, and the raw datasets resulting from the query implementation across the four queried databases.

\section*{Acknowledgments}
This study was funded by Software Center.

\section*{Disclosure of Interests}
The authors have no competing interests to declare that are relevant to the content of this article.

\bibliographystyle{splncs04}
\bibliography{bib/slr_bib}

@article{wang_survey_2024,
	title = {A survey on large language model based autonomous agents},
	volume = {18},
	issn = {2095-2236},
	url = {https://doi.org/10.1007/s11704-024-40231-1},
	language = {en},
	number = {6},
	urldate = {2026-01-27},
	journal = {Frontiers of Computer Science},
	author = {Wang, Lei and Ma, Chen and Feng, Xueyang and others},
	month = mar,
	year = {2024},
	pages = {186345},
}

@inproceedings{raheem_agentic_2025,
	title = {Agentic {AI} {Systems}: {Opportunities}, {Challenges}, and {Trustworthiness}},
	issn = {2154-0373},
	shorttitle = {Agentic {AI} {Systems}},
	url = {https://ieeexplore.ieee.org/abstract/document/11103638},
	urldate = {2026-01-27},
	booktitle = {2025 {IEEE} {International} {Conference} on {Electro} {Information} {Technology} ({eIT})},
	author = {Raheem, Tayiba and Hossain, Gahangir},
	month = may,
	year = {2025},
	pages = {618--624},
}

@article{sapkota_ai_2026,
	title = {{AI} {Agents} vs. {Agentic} {AI}: {A} {Conceptual} {Taxonomy}, {Applications} and {Challenges}},
	volume = {126},
	issn = {15662535},
	shorttitle = {{AI} {Agents} vs. {Agentic} {AI}},
	url = {http://arxiv.org/abs/2505.10468},
	urldate = {2026-01-27},
	journal = {Information Fusion},
	author = {Sapkota, Ranjan and Roumeliotis, Konstantinos I. and Karkee, Manoj},
	month = feb,
	year = {2026},
	pages = {103599},
}

@misc{peng_impact_2023,
	title = {The {Impact} of {AI} on {Developer} {Productivity}: {Evidence} from {GitHub} {Copilot}},
	shorttitle = {The {Impact} of {AI} on {Developer} {Productivity}},
	url = {http://arxiv.org/abs/2302.06590},
	urldate = {2026-01-27},
	publisher = {arXiv},
	author = {Peng, Sida and Kalliamvakou, Eirini and Cihon, Peter and Demirer, Mert},
	month = feb,
	year = {2023},
}

@article{acharya_agentic_2025,
	title = {Agentic {AI}: {Autonomous} {Intelligence} for {Complex} {Goals}—{A} {Comprehensive} {Survey}},
	volume = {13},
	issn = {2169-3536},
	shorttitle = {Agentic {AI}},
	url = {https://ieeexplore.ieee.org/abstract/document/10849561},
	urldate = {2026-01-27},
	journal = {IEEE Access},
	author = {Acharya, Deepak Bhaskar and Kuppan, Karthigeyan and Divya, B.},
	year = {2025},
	pages = {18912--18936},
}

@article{hosseini_role_2025,
	title = {The role of agentic {AI} in shaping a smart future: {A} systematic review},
	volume = {26},
	issn = {2590-0056},
	shorttitle = {The role of agentic {AI} in shaping a smart future},
	url = {https://www.sciencedirect.com/science/article/pii/S2590005625000268},
	urldate = {2026-01-27},
	journal = {Array},
	author = {Hosseini, Soodeh and Seilani, Hossein},
	month = jul,
	year = {2025},
	pages = {100399},
}

@article{denyer_developing_2008,
	title = {Developing {Design} {Propositions} through {Research} {Synthesis}},
	volume = {29},
	issn = {0170-8406},
	url = {https://doi.org/10.1177/0170840607088020},
	language = {EN},
	number = {3},
	urldate = {2026-01-27},
	journal = {Organization Studies},
	publisher = {SAGE Publications Ltd},
	author = {Denyer, David and Tranfield, David and van Aken, Joan Ernst},
	month = mar,
	year = {2008},
	pages = {393--413},
}

@article{kitchenham_guidelines_2007,
  title={Guidelines for performing systematic literature reviews in software engineering},
  author={Kitchenham, Barbara and Charters, Stuart and others},
  year={2007},
  journal={Keele}
}

@article{gough_introduction_2017,
  title={An introduction to systematic reviews},
  author={Gough, David and Thomas, James and Oliver, Sandy},
  year={2017},
  journal={SAGE}
}

@article{gusenbauer_which_2020,
	title = {Which academic search systems are suitable for systematic reviews or meta-analyses?},
	volume = {11},
	copyright = {© 2019 The Authors. Research Synthesis Methods published by John Wiley \& Sons Ltd},
	issn = {1759-2887},
	shorttitle = {Which academic search systems are suitable for systematic reviews or meta-analyses?},
	url = {https://onlinelibrary.wiley.com/doi/abs/10.1002/jrsm.1378},
	language = {en},
	number = {2},
	urldate = {2026-01-27},
	journal = {Research Synthesis Methods},
	author = {Gusenbauer, Michael and Haddaway, Neal R.},
	year = {2020},
	pages = {181--217},
}

@misc{schneider_generative_2025,
	title = {Generative to {Agentic} {AI}: {Survey}, {Conceptualization}, and {Challenges}},
	shorttitle = {Generative to {Agentic} {AI}},
	url = {http://arxiv.org/abs/2504.18875},
	urldate = {2026-01-27},
	publisher = {arXiv},
	author = {Schneider, Johannes},
	month = apr,
	year = {2025},
}

@article{murali_ai-assisted_2024,
	title = {{AI}-{Assisted} {Code} {Authoring} at {Scale}: {Fine}-{Tuning}, {Deploying}, and {Mixed} {Methods} {Evaluation}},
	volume = {1},
	shorttitle = {{AI}-{Assisted} {Code} {Authoring} at {Scale}},
	url = {https://dl.acm.org/doi/10.1145/3643774},
	number = {FSE},
	urldate = {2026-01-27},
	journal = {Proc. ACM Softw. Eng.},
	author = {Murali, Vijayaraghavan and Maddila, Chandra and Ahmad, Imad and others},
	month = jul,
	year = {2024},
	pages = {48:1066--48:1085},
}

@misc{becker_measuring_2025,
	title = {Measuring the {Impact} of {Early}-2025 {AI} on {Experienced} {Open}-{Source} {Developer} {Productivity}},
	url = {http://arxiv.org/abs/2507.09089},
	urldate = {2026-01-27},
	publisher = {arXiv},
	author = {Becker, Joel and Rush, Nate and Barnes, Elizabeth and Rein, David},
	month = jul,
	year = {2025},
}

@misc{yetistiren_evaluating_2023,
	title = {Evaluating the {Code} {Quality} of {AI}-{Assisted} {Code} {Generation} {Tools}: {An} {Empirical} {Study} on {GitHub} {Copilot}, {Amazon} {CodeWhisperer}, and {ChatGPT}},
	shorttitle = {Evaluating the {Code} {Quality} of {AI}-{Assisted} {Code} {Generation} {Tools}},
	url = {http://arxiv.org/abs/2304.10778},
	urldate = {2026-01-28},
	publisher = {arXiv},
	author = {Yetiştiren, Burak and Özsoy, Işık and Ayerdem, Miray and Tüzün, Eray},
	month = oct,
	year = {2023},
}

@article{otoum_methods_2026,
	title = {Methods and {Techniques} of {Agentic} {Software} {Engineering}: {A} {Systematic} {Literature} {Review}},
	volume = {14},
	issn = {2169-3536},
	shorttitle = {Methods and {Techniques} of {Agentic} {Software} {Engineering}},
	url = {https://ieeexplore.ieee.org/abstract/document/11343819},
	urldate = {2026-01-28},
	journal = {IEEE Access},
	author = {Otoum, Nesreen and Elkhalili, Nuha},
	year = {2026},
	pages = {7443--7465},
}

@article{bandi_rise_2025,
	title = {The {Rise} of {Agentic} {AI}: {A} {Review} of {Definitions}, {Frameworks}, {Architectures}, {Applications}, {Evaluation} {Metrics}, and {Challenges}},
	volume = {17},
	copyright = {http://creativecommons.org/licenses/by/3.0/},
	issn = {1999-5903},
	shorttitle = {The {Rise} of {Agentic} {AI}},
	url = {https://www.mdpi.com/1999-5903/17/9/404},
	language = {en},
	number = {9},
	urldate = {2026-01-28},
	journal = {Future Internet},
	publisher = {publisher},
	author = {Bandi, Ajay and Kongari, Bhavani and Naguru, Roshini and others},
	month = sep,
	year = {2025},
}

@misc{akbar_agentic_2025,
	address = {Rochester, NY},
	type = {{SSRN} {Scholarly} {Paper}},
	title = {Agentic {AI} in {Software} {Engineering}: {Practitioner} {Perspectives} {Across} the {Software} {Development} {Life} {Cycle}},
	shorttitle = {Agentic {AI} in {Software} {Engineering}},
	url = {https://papers.ssrn.com/abstract=5520159},
	language = {en},
	urldate = {2026-01-28},
	publisher = {Social Science Research Network},
	author = {Akbar, Muhammad Azeem and Khan, Arif Ali and Hamza, Muhammad and others},
	month = sep,
	year = {2025},
}

@misc{rouzrokh_lattereview_2025,
	title = {{LatteReview}: {A} {Multi}-{Agent} {Framework} for {Systematic} {Review} {Automation} {Using} {Large} {Language} {Models}},
	shorttitle = {{LatteReview}},
	url = {http://arxiv.org/abs/2501.05468},
	urldate = {2026-01-28},
	publisher = {arXiv},
	author = {Rouzrokh, Pouria and Khosravi, Bardia and Rouzrokh, Parsa and Shariatnia, Moein},
	month = oct,
	year = {2025},
}

@article{ouzzani_rayyanweb_2016,
	title = {Rayyan—a web and mobile app for systematic reviews},
	volume = {5},
	issn = {2046-4053},
	url = {https://doi.org/10.1186/s13643-016-0384-4},
	language = {en},
	number = {1},
	urldate = {2026-01-28},
	journal = {Systematic Reviews},
	author = {Ouzzani, Mourad and Hammady, Hossam and Fedorowicz, Zbys and Elmagarmid, Ahmed},
	month = dec,
	year = {2016},
	pages = {210},
}

@inproceedings{wallace_deploying_2012,
	address = {New York, NY, USA},
	series = {{IHI} '12},
	title = {Deploying an interactive machine learning system in an evidence-based practice center: abstrackr},
	isbn = {978-1-4503-0781-9},
	shorttitle = {Deploying an interactive machine learning system in an evidence-based practice center},
	url = {https://dl.acm.org/doi/10.1145/2110363.2110464},
	urldate = {2026-01-28},
	booktitle = {Proceedings of the 2nd {ACM} {SIGHIT} {International} {Health} {Informatics} {Symposium}},
	publisher = {Association for Computing Machinery},
	author = {Wallace, Byron C. and Small, Kevin and Brodley, Carla E. and others},
	month = jan,
	year = {2012},
	pages = {819--824},
}

@misc{liu_large_2025,
	title = {Large {Language} {Model}-{Based} {Agents} for {Software} {Engineering}: {A} {Survey}},
	shorttitle = {Large {Language} {Model}-{Based} {Agents} for {Software} {Engineering}},
	url = {http://arxiv.org/abs/2409.02977},
	urldate = {2026-04-27},
	publisher = {arXiv},
	author = {Liu, Junwei and Wang, Kaixin and Chen, Yixuan and others},
	month = dec,
	year = {2025},
}

@misc{he_llm-based_2025,
	title = {{LLM}-{Based} {Multi}-{Agent} {Systems} for {Software} {Engineering}: {Literature} {Review}, {Vision} and the {Road} {Ahead}},
	shorttitle = {{LLM}-{Based} {Multi}-{Agent} {Systems} for {Software} {Engineering}},
	url = {http://arxiv.org/abs/2404.04834},
	urldate = {2026-04-27},
	publisher = {arXiv},
	author = {He, Junda and Treude, Christoph and Lo, David},
	month = jul,
	year = {2025},
}

@misc{wang_agents_2024,
	title = {Agents in {Software} {Engineering}: {Survey}, {Landscape}, and {Vision}},
	shorttitle = {Agents in {Software} {Engineering}},
	url = {http://arxiv.org/abs/2409.09030},
	urldate = {2026-04-27},
	publisher = {arXiv},
	author = {Wang, Yanlin and Zhong, Wanjun and Huang, Yanxian and others},
	month = sep,
	year = {2024},
}

@misc{khoee_gonogo_2024,
	title = {{GoNoGo}: {An} {Efficient} {LLM}-based {Multi}-{Agent} {System} for {Streamlining} {Automotive} {Software} {Release} {Decision}-{Making}},
	shorttitle = {{GoNoGo}},
	url = {http://arxiv.org/abs/2408.09785},
	urldate = {2026-04-28},
	publisher = {arXiv},
	author = {Khoee, Arsham Gholamzadeh and Yu, Yinan and Feldt, Robert and others},
	month = sep,
	year = {2024},
}

@misc{kohl_automated_2026,
	title = {Automated structural testing of {LLM}-based agents: methods, framework, and case studies},
	shorttitle = {Automated structural testing of {LLM}-based agents},
	url = {http://arxiv.org/abs/2601.18827},
	urldate = {2026-04-28},
	publisher = {arXiv},
	author = {Kohl, Jens and Kruse, Otto and Mostafa, Youssef and others},
	month = jan,
	year = {2026},
}

@misc{hu_self-evolving_2024,
	title = {Self-{Evolving} {Multi}-{Agent} {Collaboration} {Networks} for {Software} {Development}},
	url = {http://arxiv.org/abs/2410.16946},
	urldate = {2026-04-28},
	publisher = {arXiv},
	author = {Hu, Yue and Cai, Yuzhu and Du, Yaxin and others},
	month = oct,
	year = {2024},
}

@article{ji_lemad_2025,
	title = {{LEMAD}: {LLM}-{Empowered} {Multi}-{Agent} {System} for {Anomaly} {Detection} in {Power} {Grid} {Services}},
	volume = {14},
	copyright = {http://creativecommons.org/licenses/by/3.0/},
	issn = {2079-9292},
	shorttitle = {{LEMAD}},
	url = {https://www.mdpi.com/2079-9292/14/15/3008},
	language = {en},
	number = {15},
	urldate = {2026-04-28},
	journal = {Electronics},
	publisher = {Multidisciplinary Digital Publishing Institute},
	author = {Ji, Xin and Zhang, Le and Zhang, Wenya and others},
	month = jan,
	year = {2025},
	pages = {3008},
}

@article{qin_soapfl_2025,
	title = {{SoapFL}: {A} {Standard} {Operating} {Procedure} for {LLM}-{Based} {Method}-{Level} {Fault} {Localization}},
	volume = {51},
	issn = {1939-3520},
	shorttitle = {{SoapFL}},
	url = {https://ieeexplore.ieee.org/document/10891926},
	number = {4},
	urldate = {2026-04-28},
	journal = {IEEE Transactions on Software Engineering},
	author = {Qin, Yihao and Wang, Shangwen and Lou, Yiling and others},
	month = apr,
	year = {2025},
	pages = {1173--1187},
}

@misc{jin_rgd_2024,
	title = {{RGD}: {Multi}-{LLM} {Based} {Agent} {Debugger} via {Refinement} and {Generation} {Guidance}},
	shorttitle = {{RGD}},
	url = {http://arxiv.org/abs/2410.01242},
	urldate = {2026-04-28},
	publisher = {arXiv},
	author = {Jin, Haolin and Sun, Zechao and Chen, Huaming},
	month = oct,
	year = {2024},
}

@inproceedings{mavani_codebase_2026,
	title = {Codebase {Aware} {Generative} {Agents} for the {SDLC}: {Automating} {Documentation}, {Dependency} {Analysis} and {Test} {Generation}},
	shorttitle = {Codebase {Aware} {Generative} {Agents} for the {SDLC}},
	url = {https://ieeexplore.ieee.org/document/11395666},
	urldate = {2026-04-28},
	booktitle = {2026 {IEEE} 5th {International} {Conference} on {AI} in {Cybersecurity} ({ICAIC})},
	author = {Mavani, Vatsal Kishorbhai},
	month = feb,
	year = {2026},
	pages = {1--4},
}

@misc{hariharan_agentic_2025,
	title = {Agentic {RAG} for {Software} {Testing} with {Hybrid} {Vector}-{Graph} and {Multi}-{Agent} {Orchestration}},
	url = {http://arxiv.org/abs/2510.10824},
	urldate = {2026-04-28},
	publisher = {arXiv},
	author = {Hariharan, Mohanakrishnan and Arvapalli, Satish and Barma, Seshu and Sheela, Evangeline},
	month = oct,
	year = {2025},
}

@inproceedings{tahat2025agree,
  title={AGREE-Dog Copilot: a neuro-symbolic approach to enhanced model-based systems engineering},
  author={Tahat, Amer and Amundson, Isaac and Hardin, David and Cofer, Darren},
  booktitle={International Conference on Bridging the Gap between AI and Reality},
  pages={117--137},
  year={2025},
  organization={Springer}
}

@inproceedings{adapa2025multi,
  title={A Multi-Agent AI Framework for Agile Workflow Automation, Issue Resolution, and Developer Performance Evaluation},
  author={Adapa, Chathurya and Anjana, ARK and Rahim, Rafsal and Victor, Ajay},
  booktitle={2025 IEEE International Conference for Women in Innovation, Technology \& Entrepreneurship (ICWITE)},
  pages={1--6},
  year={2025},
  organization={IEEE}
}

\end{document}